\documentclass[11pt]{article}

\usepackage[final]{acl}

\usepackage{times}
\usepackage{latexsym}
\usepackage{amsmath}
\usepackage{balance}
\usepackage[T1]{fontenc}

\usepackage[utf8]{inputenc}

\usepackage{microtype}
\usepackage{inconsolata}

\usepackage{graphicx}

\usepackage{xcolor}
\usepackage{url}
\usepackage{hyperref}
\hypersetup{
  colorlinks=true,
  linkcolor=blue,
  citecolor=blue,
  urlcolor=blue
}
\usepackage{algorithm}
\usepackage{algorithmic}

\title{WSDM Cup 2026 Multilingual Retrieval: A Low-Cost Multi-Stage Retrieval Pipeline}
\author{Chentong Hao\\
  Brown University\\
  Providence, RI, USA\\
  chentong\_hao@brown.edu \\\And
  Minmao Wang\\
  Fudan University\\
  Shanghai, China\\
  mmwang25@m.fudan.edu.cn \\}

\begin{document}
\maketitle

\begin{abstract}
We present a low-cost retrieval system for the WSDM Cup 2026 multilingual retrieval task, where English queries are used to retrieve relevant documents from a collection of approximately ten million news articles in Chinese, Persian, and Russian, and to output the top-1000 ranked results for each query. We follow a four-stage pipeline that combines LLM-based GRF-style query expansion with BM25 candidate retrieval, dense ranking using long-text representations from \textit{jina-embeddings-v4}, and pointwise re-ranking of the top-20 candidates using \textit{Qwen3-Reranker-4B} while preserving the dense order for the remaining results. On the official evaluation, the system achieves nDCG@20 of 0.403 and Judged@20 of 0.95. We further conduct extensive ablation experiments to quantify the contribution of each stage and to analyze the effectiveness of query expansion, dense ranking, and top-$k$ reranking under limited compute budgets.

\end{abstract}

\section{Introduction}
Multilingual retrieval aims to find relevant documents written in languages different from the user query, and is a core component of modern information access and retrieval-augmented generation systems. In this work, we present a low-cost multi-stage system for the WSDM Cup 2026 Multilingual Retrieval task\footnote{\url{https://wsdmcup-2026.github.io/}}. Given English informational queries, our goal is to retrieve relevant news documents from a large multilingual collection of approximately 10 million articles in Chinese, Persian, and Russian, and to output a ranked list of 1,000 document IDs in the standard TREC~\cite{TREC} run format. To reduce the query--document language mismatch while still performing multilingual retrieval, we index and search the machine-translated English version of each non-English document for matching and scoring, and output the corresponding original multilingual document IDs required by the task. We develop and validate our approach using the provided development queries with relevance judgments under the competition constraints. Our system is designed for rapid iteration under limited compute budgets. The core insight is that a lightweight LLM-based GRF-style expansion substantially strengthens sparse retrieval, long-context dense embeddings are critical for early precision, and restricting expensive re-ranking to a small top-$k$ (here $k=20$) captures most of the effectiveness gains on nDCG@20.

\section{Preliminary}
Let $q$ denote an English query.
Each original multilingual document is denoted by $d \in \mathcal{D}$ with a unique document ID, and its English translation text is denoted by $\tilde d$.
All retrieval and scoring stages operate on $\tilde d$, while the final output is an ordered list of 1{,}000 original document IDs in the standard TREC run format.

\section{Methodology}
\label{sec:method}

We adopt a low-cost multi-stage pipeline for multilingual news retrieval with English queries.
To unify the query-document semantics and reduce cross-language mismatch, we perform retrieval on an English rendering of the corpus by using the machine-translated English text for each non-English document.
We use the English query to match and score these translated documents, and output the corresponding original multilingual document IDs required by the task.
The system consists of four stages: (1) GRF-style~\cite{GRF} query expansion with an LLM, (2) sparse candidate generation with BM25~\cite{BM25}, (3) dense ranking with \textit{jina-embeddings-v4}~\cite{JinaV4}, and (4) top-$k$ re-ranking with \textit{Qwen3-Reranker-4B}~\cite{Qwen3-Embedding}.

\subsection{Stage 1: GRF-style query expansion}

Given an English query $q$, we prompt \textit{deepseek-chat} to generate a short news-style pseudo-document $g(q)$.
We preprocess $g(q)$ with spaCy \texttt{en\_core\_web\_sm} (lemmatization + stopword removal), rank tokens by term frequency, filter tokens also appearing in the preprocessed query, and append the top-30 remaining terms to $q$.

\paragraph{GRF-style expansion definition.}
We use the prompt template in Appendix~\ref{app:prompt} and sample with temperature=0.7, top\_p=1.0, max\_tokens=512.

\paragraph{Term extraction.}
Let $\mathcal{T}(g(q))$ be the token multiset after preprocessing.
We compute $\mathrm{tf}(t; g(q))$, remove tokens in the preprocessed query, and keep the top $\theta$ terms $E=\{e_1,\dots,e_\theta\}$.
The expanded query is:
\begin{equation}
q' = q \;\Vert\; \big( e_1 \Vert e_2 \Vert \cdots \Vert e_\theta \big),
\label{eq:qexpand}
\end{equation}
where $\Vert$ denotes string concatenation with whitespace.

\begin{algorithm}[t]
\caption{GRF-style Query Expansion}
\label{alg:grf}
\begin{algorithmic}[1]
\STATE \textbf{Input:} query $q$, expansion size $\theta$
\STATE Generate $g(q)$ with \textit{deepseek-chat} 
\STATE Preprocess $g(q)$ and $q$ with spaCy (lemmatize + remove stopwords)
\STATE Select top $\theta$ terms $E=\{e_1,\dots,e_\theta\}$ by $\mathrm{tf}(\cdot; g(q))$, filtering terms in the preprocessed $q$
\STATE Output $q' = q \;\Vert\; (e_1 \Vert \cdots \Vert e_\theta)$ (Eq.~\ref{eq:qexpand})
\end{algorithmic}
\end{algorithm}

\subsection{Stage 2: Sparse retrieval}

We run BM25 over a single index built on the English translation view $\tilde d$ of the full multilingual collection.
Given the expanded query $q'$, we retrieve the top-2000 candidate documents and denote the candidate set as $\mathcal{C}_{2000}(q')$.

\paragraph{BM25 configuration.}
We use BM25 with default BM25 parameters. All candidates are mapped back to their original multilingual document IDs for subsequent stages and final output.

\subsection{Stage 3: Dense ranking}

We compute dense representations for the query and candidate documents using \textit{jina-embeddings-v4}.
For each candidate $d \in \mathcal{C}_{2000}(q')$, we embed its translation text $\tilde d$ and rank candidates by cosine similarity to the query embedding.
This produces a dense top-1000 list per query.

\paragraph{Document input construction.}
We construct the embedding input from the translated document text $\tilde d$ by concatenating available fields such as title and body.

\subsection{Stage 4: Top-$k$ re-ranking}

We re-rank only the top $k=20$ candidates using \textit{Qwen3-Reranker-4B}.
For each query--document pair, we construct an instruction-formatted input and compute relevance as $P(\texttt{yes})$ from the last-token logits for the $\{\texttt{yes}, \texttt{no}\}$ decision.
The re-ranked top-20 documents are placed first, and documents ranked 21--1000 preserve their original dense order.

\paragraph{Pointwise scoring.}
Let the reranker output logits over the vocabulary for the next token.
We compute a binary relevance probability by restricting to label tokens $\{\tau_{\texttt{yes}}, \tau_{\texttt{no}}\}$ and applying a 2-way softmax:
\begin{equation}
\small
P(\texttt{yes}\mid q,\tilde d) = \frac{\exp(\ell(\tau_{\texttt{yes}}))}{\exp(\ell(\tau_{\texttt{yes}}))+\exp(\ell(\tau_{\texttt{no}}))}.
\end{equation}
We use the prompt template in Appendix~\ref{app:prompt} and truncate $\tilde d$ to fit the model context window.

\section{Experiments}
\label{sec:exp}

\subsection{Task setup and evaluation}

We follow the WSDM Cup 2026 Multilingual Retrieval task setting.
Systems receive English informational queries and retrieve relevant news documents from a large multilingual collection containing Chinese, Persian, and Russian articles. For each query, the system outputs a ranked list of 1{,}000 documents in the standard TREC run format. Submissions are evaluated with nDCG@20 and Judged@20.

\subsection{Data and constraints}

The organizers provide a small set of development queries with relevance judgments for system development. Following the competition rules, we do not use any publicly available NeuCLIR relevance assessments beyond the released development labels. All retrieval and re-ranking stages operate fully automatically without human intervention. The organizers also provide an offline version of the dataset and evaluation resources. We conduct our ablation experiments on this offline setup, which enables convenient repeated runs and logging for analysis compared with the online submission pipeline.

\subsection{Baselines}
\label{sec:baseline}

We compare our submission against a set of organizer-reported baseline runs that span sparse retrieval, dense embedding retrieval, and LLM-based retrieval models under the same WSDM Cup 2026 evaluation setting. We include \textit{BGE-M3 Sparse} \citep{BGE-M3}, \textit{e5 Large} \citep{E5-Large}, \textit{RepLlama} \citep{RepLlama}, and \textit{Qwen3-0.6B-Embed} \citep{Qwen3-Embedding}. We also report a strong sparse baseline \textit{BM25} \citep{BM25} and a recent high-quality embedding model \textit{Arctic-Embed Large v2} \citep{Arctic-Embed-Large-v2}. Finally, we include multilingual retrieval systems \textit{JinaV3} \citep{JinaV3}, \textit{MILCO} \citep{MILCO}, and \textit{PLAID-X} \citep{PLAID-X}, which achieve the best effectiveness among the baselines in our comparison.

\subsection{Implementation details}
Our pipeline consists of GRF-style query expansion, sparse retrieval, dense ranking, and top-$k$ re-ranking (Section~\ref{sec:method}).
For GRF expansion, we generate a news-style pseudo-document with \textit{deepseek-chat} (temperature=0.7, top\_p=1.0, max\_tokens=512), extract the top $\theta=30$ expansion terms after lemmatization and stopword filtering, and append them to the original query.
Sparse retrieval uses BM25 to retrieve top-2000 candidates per query.
Dense ranking uses \textit{jina-embeddings-v4} with truncate\_dim=1024 and max\_length=5120, and ranks candidates by cosine similarity after L2-normalization.
Finally, we re-rank only the top $k=20$ documents using \textit{Qwen3-Reranker-4B} with a pointwise \texttt{yes/no} scoring prompt.

\subsection{Ablation setting}

\begin{itemize}
  \item \textbf{BM25}: Sparse retrieval using the base query $q$ only, with no dense ranking.
  \item \textbf{BM25 with GRF}: Sparse retrieval using the expanded query $q'$ only, with no dense ranking.
  \item \textbf{Jina + BM25}: Candidates are retrieved by BM25 using the base query $q$ (top-2000), then \textit{jina-embeddings-v4} rank candidates by cosine similarity to produce the top-1000 list.
  \item \textbf{Jina with GRF + BM25}: Candidates are retrieved by BM25 using the base query $q$ (top-2000), then the dense stage uses \textit{jina-embeddings-v4} with an expanded query representation (based on $q'$) to rank candidates by cosine similarity.
  \item \textbf{Jina with GRF + BM25 with GRF}: Candidates are retrieved by BM25 using the expanded query $q'$ (top-2000), then \textit{jina-embeddings-v4} rank candidates by cosine similarity to produce the top-1000 list.
\end{itemize}


\subsection{Main results}

Table~\ref{tab:main} reports the official evaluation results of our final run on the WSDM Cup 2026 test set, and compares our submission against a subset of organizer-reported baseline systems with lower nDCG@20. Our approach achieves an nDCG@20 of 0.403 with high Judged@20 of 0.95, indicating that most of the top-ranked documents are within the judged pool while maintaining competitive early precision.
Overall, the results suggest that combining lightweight LLM-based query expansion with sparse candidate generation, dense retrieval, and a small top-$k$ reranking stage can yield strong effectiveness under limited compute budgets.

\begin{table}[t]
\centering

\resizebox{\linewidth}{!}{%
\begin{tabular}{lcc}

\hline
Model & nDCG@20 & Judged@20 \\
\hline
BGE-M3 Sparse & 0.054 & 0.30 \\
e5 Large & 0.205 & 0.70 \\
RepLlama & 0.255 & 0.86 \\
Qwen3 0.6B Embed & 0.309 & 0.92 \\
BM25 & 0.349 & \textbf{0.97} \\
Arctic-Embed Large v2 & 0.352 & 0.92 \\
JinaV3 & 0.355 & 0.93 \\
MILCO & \underline{0.395}& 0.93 \\
\hline
Ours & \textbf{0.403} & \underline{0.95} \\
\hline
\end{tabular}}
\caption{Comparison against selected organizer-reported baseline runs. Best baseline is in bold and second-best baseline is underlined.}

\label{tab:main}
\end{table}

\subsection{Ablation results}

Table~\ref{tab:ablation} presents an ablation study of the main components in our pipeline. Adding GRF-style query expansion to sparse BM25 retrieval substantially improves early ranking quality, increasing nDCG@20 from 0.3306 to 0.4020, while also improving recall (R@1000 from 0.5867 to 0.6526). Replacing sparse-only ranking with dense ranking using \textit{jina-embeddings-v4} yields a large additional gain (nDCG@20 = 0.4975), highlighting the importance of semantic representations for promoting relevant documents to the top ranks. Incorporating GRF on top of dense ranking provides a consistent but smaller improvement. The best-performing setting combines GRF-based expansion with dense ranking and achieves the highest nDCG@20 of 0.5158. Notably, this setting slightly reduces R@1000 compared to dense ranking over BM25-base candidates, suggesting a precision--recall trade-off: expanded queries can shift the candidate distribution by injecting high-specificity entities/events that promote near-duplicate topical matches, improving early precision but reducing diversity within the top-1000.

\begin{table}[t]
\centering
\resizebox{\linewidth}{!}{%
\begin{tabular}{lcc}
\hline
Setting & R@1000 & nDCG@20 \\
\hline
BM25 & 0.5867 & 0.3306 \\
BM25 with GRF & 0.6526 & 0.4020 \\
Jina + BM25 & 0.6637 & 0.4975 \\
Jina with GRF + BM25 & \textbf{0.6656} & 0.5015 \\
Jina with GRF + BM25 with GRF & 0.6330 & \textbf{0.5158} \\
\hline
\end{tabular}}
\caption{Ablation results on the development set, focusing on the impact of GRF-style query expansion and dense ranking. We report recall at 1,000 and nDCG@20.}
\label{tab:ablation}
\end{table}

\subsection{Efficiency considerations}

Our design emphasizes rapid iteration under limited compute budgets.
In particular, we restrict expensive LLM re-ranking to the top $k=20$ documents per query, while preserving the remaining ranks from the dense stage.
This reduces inference cost while still improving early precision measured by nDCG@20.

\section{Conclusion}
\label{sec:conclusion}

We presented a low-cost, fully automated multi-stage multilingual retrieval pipeline for the WSDM Cup 2026 task.
The system combines LLM-based GRF-style query expansion, BM25 sparse candidate generation, long-text dense ranking with \textit{jina-embeddings-v4}, and lightweight top-$k$ re-ranking using \textit{Qwen3-Reranker-4B}. By restricting expensive re-ranking to only the top 20 documents while preserving the remaining dense ranking order, the approach supports rapid iteration under limited compute budgets and yields strong early-ranking effectiveness as reflected by the official nDCG@20 results. Future work may explore (i) more effective query rewriting strategies beyond GRF-style expansion, (ii) training cross-lingual retrieval models that jointly optimize English queries and multilingual documents (or their translations) to better handle language mismatch, and (iii) multi-agent retrieval architectures where specialized agents collaborate (e.g., rewriting, sparse retrieval, dense retrieval, and reranking) to improve robustness and efficiency.

\appendix
\section{Prompts and scoring templates}
\label{app:prompt}

\paragraph{GRF pseudo-document generation prompt.}

\begin{quote}\small
You are a professional news writer.
Based on the query, write a short, factual news-style article (3–6 paragraphs).

Query: \{QUERY\}

Write the article:
\end{quote}

\paragraph{Qwen reranker prompt template.}
\begin{quote}\small
You are a search relevance judge. Given an instruction, a query, and a candidate news document, decide whether the document satisfies the instruction and is relevant to the query.
Answer with only one token: yes or no.

Instruct: Given a news search query, retrieve relevant news articles that answer the query.

Query: \{QUERY\}

Document: \{DOC\}

Answer:
\end{quote}

\bibliography{acl_latex}
\balance

\end{document}